\documentclass[runningheads]{svmult}

\usepackage{makeidx}

\usepackage{graphicx}

\usepackage{subeqnar}

\usepackage{multicol}

\usepackage{cropmark}

\usepackage{physprbb}

\def\leti{Lense--Thirring}
\def\zone{the error due to the even zonal harmonics of geopotential\ }
\def\bm#1{{\mbox{\boldmath$#1$\unboldmath}}}
\def\rfr#1{(\ref{#1})}

\def\eqi{\begin{equation}}
\def\eqf{\end{equation}}
\def\eqia{\begin{eqnarray}}
\def\eqfa{\end{eqnarray}}

\def\lb#1{\label{#1}}

\begin{document}

\title*{The impact of the new CHAMP and GRACE Earth gravity models on
the measurement of the general relativistic Lense--Thirring effect
with the LAGEOS and LAGEOS II satellites }

\toctitle{ The impact of the new CHAMP and GRACE Earth gravity
models\protect\newline on the measurement of the general
relativistic Lense--Thirring effect}

\titlerunning{The impact of the new CHAMP and GRACE Earth gravity models}

\author{Lorenzo Iorio\inst{1} }

\authorrunning{Lorenzo Iorio}

\institute{Dipartimento di Fisica dell'Universit${\rm \grave{a}}$
di Bari, via Amendola 173, 70126, Bari, Italy}

\maketitle

\begin{abstract}
Among the effects predicted by the General Theory of Relativity
for the orbital motion of a test particle, the post-Newtonian
gravitomagnetic Lense-Thirring effect is very interesting and, up
to now, there is not yet an undisputable direct experimental test
of it. To date, the data analysis of the orbits of the existing
geodetic LAGEOS and LAGEOS II satellites has yielded a test of the
Lense-Thirring effect with a claimed accuracy of 20$\%$-30$\%$.
According to some scientists such estimates could be optimistic.
Here we wish to discuss the improvements obtainable in this
measurement, in terms of reliability of the evaluation of the
systematic error and reduction of its magnitude, due to the new
CHAMP and GRACE Earth gravity models.

\end{abstract}

\section{Introduction}
The linearized weak--field and slow--motion approximation of the
General Theory of Relativity (GTR) \cite{ciuwhe95} is
characterized by the condition $g_{\mu\nu}\sim
\eta_{\mu\nu}+h_{\mu\nu}$ where $g_{\mu\nu}$ is the curved
spacetime metric tensor, $\eta_{\mu\nu}$ is the Minkowski metric
tensor of the flat spacetime of Special Relativity and the
$h_{\mu\nu}$ are small corrections such that $|h_{\mu\nu}| \ll 1$.
Until now, many of its predictions, for the motion of light rays
and test masses have been tested, in the Solar System, with a
variety of techniques to an accuracy level of the order of $0.1\%$
\cite{wil01}. It is not so for the gravitomagnetic\footnote{In the
weak field and slow motion approximation of GTR the equations of
motion of a test particle freely falling in the gravitational
field of a central spinning body are formally analogous to those
governing the motion of an electrically charged particle in an
electromagnetic field under the action of the velocity--dependent
Lorentz force. In the gravitational case the role of the magnetic
field is played by the so called gravitomagnetic field which is
generated by the off--diagonal terms of the metric $g_{0i}$ and
whose source is the proper angular momentum \bm J of the central
body.} Lense--Thirring effect due to its extreme smallness. It can
be thought of as a consequence of a gravitational spin--spin
coupling.

If we consider the motion of a spinning particle in the
gravitational field of a central body of mass $M$ and proper
angular momentum \bm J,  it turns out that the spin \bm s of the
orbiting particle undergoes a tiny precessional motion
\cite{sch60}. The most famous experiment devoted to the
measurement, among other things, of such gravitomagnetic effect in
the gravitational field of Earth is the Stanford University GP--B
mission \cite{eveetal01} which should fly at the end of 2003.

If the whole orbit of a test particle in its geodesic motion
around $M$ is considered as a sort of giant gyroscope, its orbital
angular momentum \bm \ell\ undergoes the Lense--Thirring
precession, so that the longitude of the ascending node $\Omega$
and the argument of pericentre $\omega$ of the orbit of the test
particle are affected by tiny secular precessions $\dot\Omega_{\rm
LT}$, $\dot\omega_{\rm LT}$ \cite{leti18, ciuwhe95}
\begin{equation}
\dot\Omega_{\rm LT} =\frac{2GJ}{c^2 a^3(1-e^2)^{\frac{3}{2}}},\
\dot\omega_{\rm LT} =-\frac{6GJ\cos i}{c^2
a^3(1-e^2)^{\frac{3}{2}}},
\end{equation} where $a,\ e$ and $i$ are the semimajor axis, the eccentricity and the inclination, respectively, of
the orbit, $c$ is the speed of light and $G$ is the Newtonian
gravitational constant.

Up to now, the only attempts to detect the \leti\ effect on the
orbit of test particles in the gravitational field of Earth are
due to Ciufolini and coworkers \cite{ciuf00} who analysed the
laser data of the existing LAGEOS and LAGEOS II satellites over
time spans of some years. The observable is a suitable combination
of the orbital residuals of the nodes of LAGEOS and LAGEOS II and
the perigee\footnote{The perigee of LAGEOS is not a good
observable because $e_{\rm LAGEOS}=0.0045$, while $e_{\rm LAGEOS\
II }=0.014$.} of LAGEOS II according to an idea exposed in
\cite{ciuf96} \eqi\delta\dot\Omega^{\rm L
}+c_1\delta\dot\Omega^{\rm L\ II}+c_2\delta\dot\omega^{\rm L\
II}\sim 60.2\mu_{\rm LT},\ c_1\sim 0.295,\ c_2\sim
-0.35,\lb{ciufform}\eqf where the superscripts I and II refer to
LAGEOS and LAGEOS II, respectively. The quantity $\mu_{\rm LT}$ is
the solved--for least square parameter which is 0 in Newtonian
mechanics and 1 in GTR. The Lense-Thirring signature, entirely
adsorbed in the residuals of $\dot\Omega$ and $\dot\omega$ because
the gravitomagnetic force has been purposely set equal to zero in
the force models, is a linear trend with a slope of 60.2
milliarcseconds per year (mas yr$^{-1}$ in the following). The
standard, statistical error is evaluated as 2$\%$. The claimed
total accuracy, including various sources of systematic errors, is
of the order of $20\%-30\%$.

The main sources of systematic errors in this experiment are
\begin{itemize}
\item
the unavoidable aliasing effect due to the mismodelling in the
classical secular precessions induced on $\Omega$ and $\omega$ by
the even zonal coefficients $J_{l}$ of the multipolar expansion of
geopotential
\item
the non--gravitational perturbations affecting especially the
perigee of LAGEOS II \cite{luc01, luc02}. Their impact on the
proposed measurement is difficult to be reliably assessed
\cite{riesetal98}
\end{itemize}
It turns out that the mismodelled classical precessions due to the
first two even zonal harmonics of geopotential $J_2$ and $J_4$ are
the most insidious source of error for the Lense--Thirring
measurement with LAGEOS and LAGEOS II. The combination
\rfr{ciufform} is insensitive just to $J_2$ and $J_4$. According
to the full covariance matrix of the EGM96 gravity model
\cite{lemetal98}, the error due to the remaining uncancelled even
zonal harmonics amounts to almost 13$\%$ \cite{iorcelmec03}.
However, if the correlations among the even zonal harmonic
coefficients are neglected and the variance matrix is used in a
Root--Sum--Square fashion\footnote{Such approach is considered
more realistic by some authors \cite{riesetal98} because nothing
assures that the correlations among the even zonal harmonics of
the covariance matrix of the EGM96 model, which has been obtained
during a multidecadal time span, would be the same during an
arbitrary past or future time span of a few years as that used in
the LAGEOS--LAGEOS II
\leti\ experiment.},
\zone amounts to 46.6$\%$ \cite{iorcelmec03}. With this estimate
and the evaluations of \cite{luc01, luc02} for the impact of the
non--gravitational perturbations the total error in the
LAGEOS--LAGEOS II
\leti\ experiment would be of the order of 50$\%$. If the sum of
the absolute values of the individual errors is assumed, an upper
bound of 83$\%$ for the systematic error due to the even zonal
harmonics of geopotential is obtained; then, the total error in
the LAGEOS--LAGEOS II
\leti\ experiment would become of the order of 100$\%$. This
evaluations agree with those released in \cite{riesetal98}.

The originally proposed LAGEOS III/LARES mission \cite{ciuf86}
consists of the launch of a LAGEOS--type satellite--the
LARES--with the same orbit of LAGEOS except for the inclination
$i$ of its orbit, which should be supplementary to that of LAGEOS,
and the eccentricity $e$, which should be one order of magnitude
larger in order to perform other tests of post--Newtonian gravity
\cite{iorciufpav02, ioryuk02}. The choice of the particular value
of the inclination for LARES is motivated by the fact that in this
way, by using as observable the sum of the nodes of LAGEOS and
LARES, it should be possible to cancel out to a very high level
all the contributions of the even zonal harmonics of
geopotential, which depends on $\cos i$, and add up the
Lense--Thirring precessions which, instead, are independent of
$i$. The use of the nodes would allow to reduce greatly the impact
of the non--gravitational perturbations to which such Keplerian
orbital elements are rather insensitive \cite{luc01, luc02}.

In \cite{iorlucciuf02} an alternative observable based on the
combination of the residuals of the nodes of LAGEOS, LAGEOS II and
LARES and the perigee of LAGEOS II and LARES has been proposed. It
would allow to cancel out the first four even zonal harmonics so
that the error due to the remaining even zonal harmonics of
geopotential would be rather insensitive both to the unavoidable
orbital injection errors in the LARES inclination and to the
correlations among the even zonal harmonic coefficients. It would
amount to $0.02\%$--$0.1\%$ only \cite{iorlucciuf02, ior03} (EGM96
full covariance and variance RSS calculations).

In regard to the present status of the LARES project,
unfortunately, up to now, although its very low cost with respect
to other much more complex and expensive space--based missions, it
has not yet been approved by any national space agency or
scientific institution.
\section{The impact of the CHAMP and GRACE Earth gravity models}
From the previous considerations it could be argued that, in order
to have a rather precise and reliable estimate of the total
systematic error in the measurement of the Lense--Thirring effect
with the existing LAGEOS satellites it would be better to reduce
the impact of geopotential in the error budget and/or discard the
perigee of LAGEOS II which is very difficult to handle and is a
relevant source of uncertainty due to its great sensitivity to
many non--gravitational perturbations.

The forthcoming more accurate Earth gravity models from CHAMP
\cite{pav00} and, especially, GRACE \cite{riesetal03} will yield
an opportunity to realize both these goals, at least to a certain
extent.

In order to evaluate quantitatively the opportunities offered by
the new terrestrial gravity models we have preliminarily used the
recently released EIGEN2 gravity model \cite{reigetal03}.

With regard to the combination \rfr{ciufform}, it turns out that
the systematic error due to the even zonal harmonics of
geopotential, according to the full covariance matrix of EIGEN2 up
to degree $l=70$, amounts to 7$\%$, while if the diagonal part
only is adopted it becomes 9$\%$ (RSS calculation). Of course,
even if the LAGEOS and LAGEOS II data had been reprocessed with
the EIGEN2 model, the problems posed by the correct evaluation of
the impact of the non--gravitational perturbations on the perigee
of LAGEOS II would still persist.

A different approach could be followed by taking the drastic
decision of canceling out only the first even zonal harmonic of
geopotential by discarding at all the perigee of LAGEOS II. The
hope is that the resulting gravitational error is reasonably small
so to get a net gain in the error budget thanks to the fact that
the nodes of LAGEOS and LAGEOS II exhibit a very good behavior
with respect to the non--gravitational perturbations. Indeed, they
are far less sensitive to their tricky features than the perigee
of LAGEOS II. Moreover, they can be easily and accurately
measured, so that also the formal, statistical error should be
reduced. A possible observable is\footnote{A similar approach is
presented in \cite{riesetal03}, although without quantitative
details.} \eqi\delta\dot\Omega^{\rm L }+c_1\delta\dot\Omega^{\rm
L\ II}\sim 48.2\mu_{\rm LT},\ c_1\sim 0.546.\lb{iorform}\eqf
According to the full covariance matrix of EIGEN2 up to degree
$l=70$, the systematic error due to the even zonal harmonics from
$l=4$ to $l=70$ amounts to 8.5 mas yr$^{-1}$ yielding a 17.8$\%$
percent error, while if the diagonal part only is adopted it
becomes\footnote{The even zonal harmonics are much more mutually
uncorrelated in EIGEN2 than in EGM96.} 22$\%$ (RSS calculation).
EGM96 would not allow to adopt \rfr{iorform} because its full
covariance matrix up to degree $l=70$ yields an error of 47.8$\%$
while the error according to its diagonal part only amounts even
to 104$\%$ (RSS calculation). Note also that the combination
\rfr{iorform} preserves one of the most important features of the
combination \rfr{ciufform}: indeed, it allows to cancel out the
very insidious 18.6-year tidal perturbation which is a $l=2,\ m=0$
constituent with a period of 18.6 years due to the Moon's node and
nominal amplitudes of the order of 10$^3$ mas on the nodes of
LAGEOS and LAGEOS II \cite{ior01}. On the other hand, the impact
of the non--gravitational perturbations on the combination
\rfr{iorform} over a time span of, say, 7 years can be quantified
in just 0.1 mas yr$^{-1}$, yielding a 0.3$\%$ percent error. The
results of Table 2 and Table 3 of \cite{iorlucciuf02} have been
used. It is also important to notice that, thanks to the fact that
the periods of many gravitational and non--gravitational
time--dependent perturbations acting on the nodes of the LAGEOS
satellites are rather short, a reanalysis of the LAGEOS and LAGEOS
II data over just a few years could be performed. This is not so
for the combination \rfr{ciufform} because some of the
gravitational \cite{ior01} and non--gravitational \cite{luc01}
perturbations affecting the perigee of LAGEOS II have periods of
many years. Then, with a little time--consuming reanalysis of the
nodes only of the existing LAGEOS and LAGEOS II satellites with
the EIGEN2 data it would at once be possible to obtain a more
accurate and reliable measurement of the Lense--Thirring effect,
avoiding the problem of the uncertainties related to the use of
the perigee of LAGEOS II.

Very recently the first preliminary Earth gravity models including
some data from GRACE have been released; among them the GGM01C
model\footnote{It can be retrieved on the WEB at
http://www.csr.utexas.edu/grace/gravity/}, which combines the
Center for Space Research (CSR) TEG4 model with data from GRACE,
seems to be very promising. Indeed, the error due to geopotential
in the combination \rfr{ciufform}, evaluated by using the variance
matrix only (RSS calculation), amounts to 2.2$\%$ (with an upper
bound of 3.1$\%$ obtained from the sum of the absolute values of
the individual terms). Instead, the combination \rfr{iorform}
would be affected at almost 14$\%$ level (RSS calculation), with
an upper bound of almost 18$\%$ from the sum of the absolute
values of the individual errors. However, it should be pointed out
that extensive calibration tests have to be still performed for
the GGM01C model.
\section{Conclusions}
When more robust and complete terrestrial gravity models from
CHAMP and GRACE will be available in the near future the
combination \rfr{iorform} could hopefully allow for a measurement
of the Lense--Thirring effect with a total systematic error,
mainly due to geopotential, of some percent over a time span of a
few years without the uncertainties related to the evaluation of
the impact of the non--gravitational perturbations acting upon the
perigee of LAGEOS II. On the other hand, the obtainable accuracy
with the combination \rfr{ciufform}, whose error due to
geopotential is smaller than that of \rfr{iorform}, is strongly
related to improvements in the evaluation of the
non--gravitational part of the error budget and to the use of time
spans of many years.


\end{document}